# Experimental and numerical measurement of the thermal performance for parabolic trough solar concentrators

**Khaled Mohamad, Philippe Ferrer**
School of Physics, University of the Witwatersrand, Johannesburg 2001, South Africa

khaled@aims.edu.gh

**Abstract** Parabolic trough mirror plants are a popular design for conversion of solar energy to electricity via thermal processes. The receiver unit (RU) for absorbing the concentrated solar radiation is limited to a maximum temperatures (580°C) and is responsible for efficiency losses mainly via thermal radiation. We built a RU in the laboratory to study the thermal performance for different designs and we companied this study with a mathematical module implemented on a simulation code. In this work, the simulation and the first set of experiments show a good agreement, validating applicability of the code.

## 1. Introduction

A parabolic trough solar thermal power plant consists of a series of parabolic mirrors concentrating solar radiation onto a linear focal line along which the receiver unit is positioned. The receiver heats up and in turn imparts a large portion of its heat to a heat transfer fluid circulating within. This heat transfer fluid can then be utilized in a steam cycle to generate electricity. The receiver is one of the most complex parts and the efficiency of the entire system is largely determined by it. It has to be carefully designed in such a way so as to minimize the energy losses. Every part of the receiver unit is a topic of ongoing research, such as the working fluid that can be used, and also the optical, chemical, and thermal properties of the concerned material [1].

Typically, the receiver unit consists of a blackened absorber pipe (AP) encapsulated by the glass cover (GC), (See Fig. (1a)). There is a vacuum in between to minimize convective heat losses [2]. For the conduction losses, the thermal contacts between the receiver pipe and the glass cover are kept to a minimum. The heat transfer fluid (HTF) inside the receiver pipe is heated by the concentrated solar radiation. The hot HTF can be used in generating electricity through a steam cycle or in thermochemical applications [3]. The dominant heat losses at high temperatures are due to the thermal emission (IR) from the receiver pipe. There is a conventional method to minimize the IR by painting the receiver pipe with a spectrally selective coating, a dielectric film that absorbs well in the visible region of the solar spectrum and emits poorly in the IR region. Much work has been published in regard to the selective coating and their properties [4]. The main weakness of selective coating is that it prevents the receiver pipe from being heated to high temperature, since it thermally decomposes at about 680 K [5], [6].

An alternative option to the selective coating is to reduce the thermal emission from the glass cover tube of a trough collector by trapping IR via a reflective surface on the part of the glass not facing the trough.

The solar radiation inlet may be coated with a hot mirror type coating [7–10], i.e., Indium-Tin-Oxide and applying the reflective cavity around the absorber [11], which is shown in Fig. (1b).

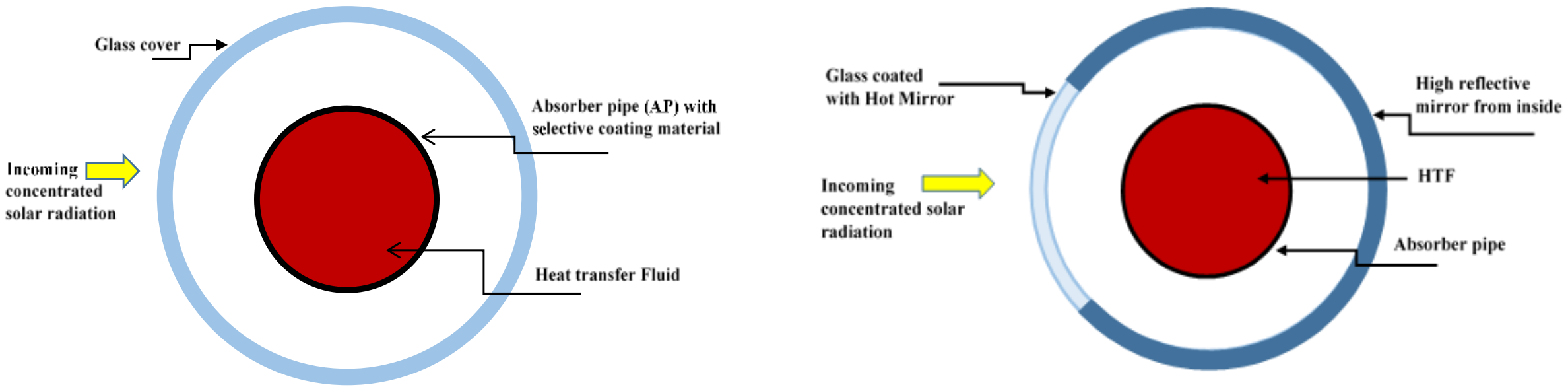

**Figure 1.** a) Receiver unit with a selective coating. b) Receiver unit with a cavity design [11].

Measuring the heat losses of the receiver unit of the parabolic trough is an effective way to determine the efficiency of the RU especially at a higher Heat transfer fluid (HTF) temperature. It allows us to judge the performance of different RU designs. In this work, we studied the thermal performance of the RU by measuring the heat losses experimentally and compared our results with our simulation code, which is based on a mathematical model of the system.

## 2. Experiment description

The Receiver Unit (RU) or heat collector element was tested indoor. Two heating elements inside the absorber pipe of the RU brought the HTF temperature to a desirable test temperature. In one experimental set, the glass tube was evacuated, in the other air was left inside the annulus. The heating element power was adjusted to a desired power value using a variac. Once the RU reached a steady state temperature, the electrical power required to maintain the HTF temperature equaled the heat loss of the RU at that temperature. Further, the temperatures along the RU elements were constant. The heat loss of the RU was tested at different temperatures corresponding to heating power generation of the heating elements from 50 W to 1.5 kW in roughly 50 W increments. Heat losses were reported per meter of RU. The temperature around the absorber pipe and the glass cover was measured with thermocouples and hence we could calculate the emittance of AP and the heat loss to the environment at different temperatures.

The length of the tested RU was 2.7 m at 25 °C (It will expand by about 9 mm at 300 °C with a mild steel absorber pipe outer/inner diameter of 3.2/2.8 cm and a two pieces of a Pyrex glass cover outer/inner diameter of 5.8/5.4 cm with a length of 1.35 m each. The Pyrex glass pieces were joined together with a brass section in the centre of the AP. We vacuum insulated the central brass piece, glass covers and the AP together using flame resistant high temperature silicon. The high temperature silicon also acted as a thermal insulator between the joints in this experiment.

The heating elements were two 1.2 m with 8 mm outer diameter. In order to prevent the heating element from touching the AP, we introduced spacers to center the heating elements. These spacers can be made from steel with a small size and sharp edge toward the AP inner surface to minimize the losses through it. Thermocouples measured the temperature of the outer surface of the AP and GC. Thermocouples were introduced to determine the average temperatures and heating behavior of the RU. The Absorber pipe and Glass cover temperatures were measured using two and four K-type thermocouples respectively. A good contact between the thermocouples and the surface was required for accurate measurement. Figs.2 and 3 show more details about the setup.
As indicated in Fig. 3, the thermocouples wires are insulated with a thermally resistance braided tube and grouped together at the end. Their ends are connected to the control and measurement unit.

Furthermore, two holes through the end piece are made to allow the AP thermocouples wires get out and connected to the control and measurement unit via vacuum tight exits. The control and measurement unit allowed us to regulate and adjust the heater power in accordance to temperature requirements.

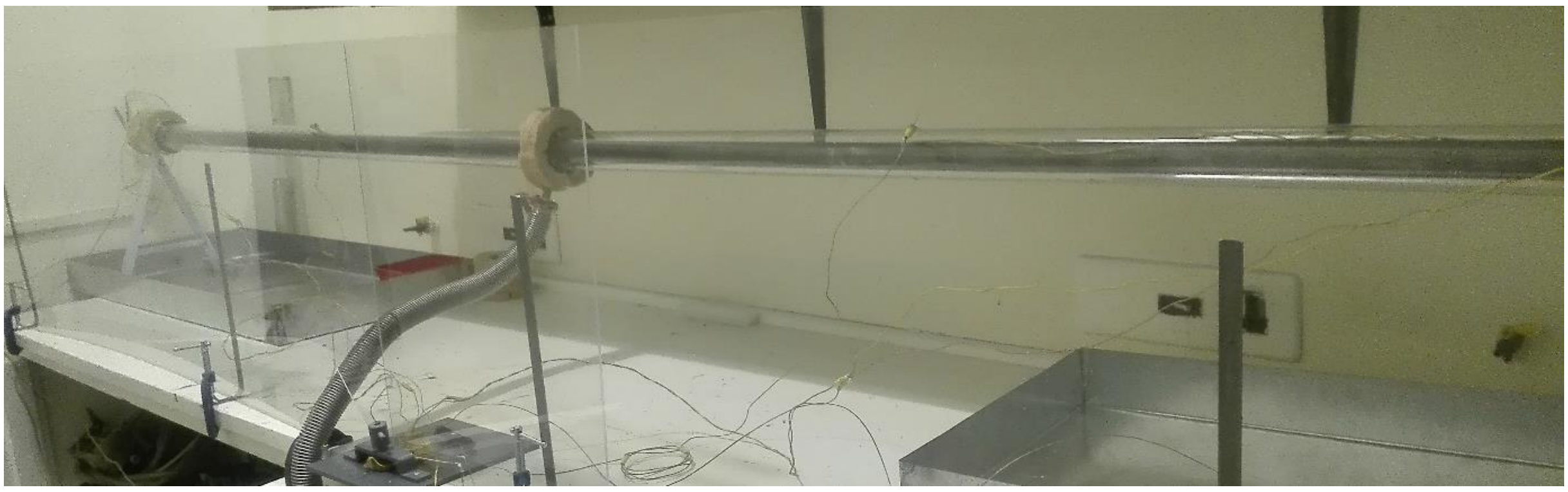

**Figure 2**: The receiver unit set up in the laboratory.

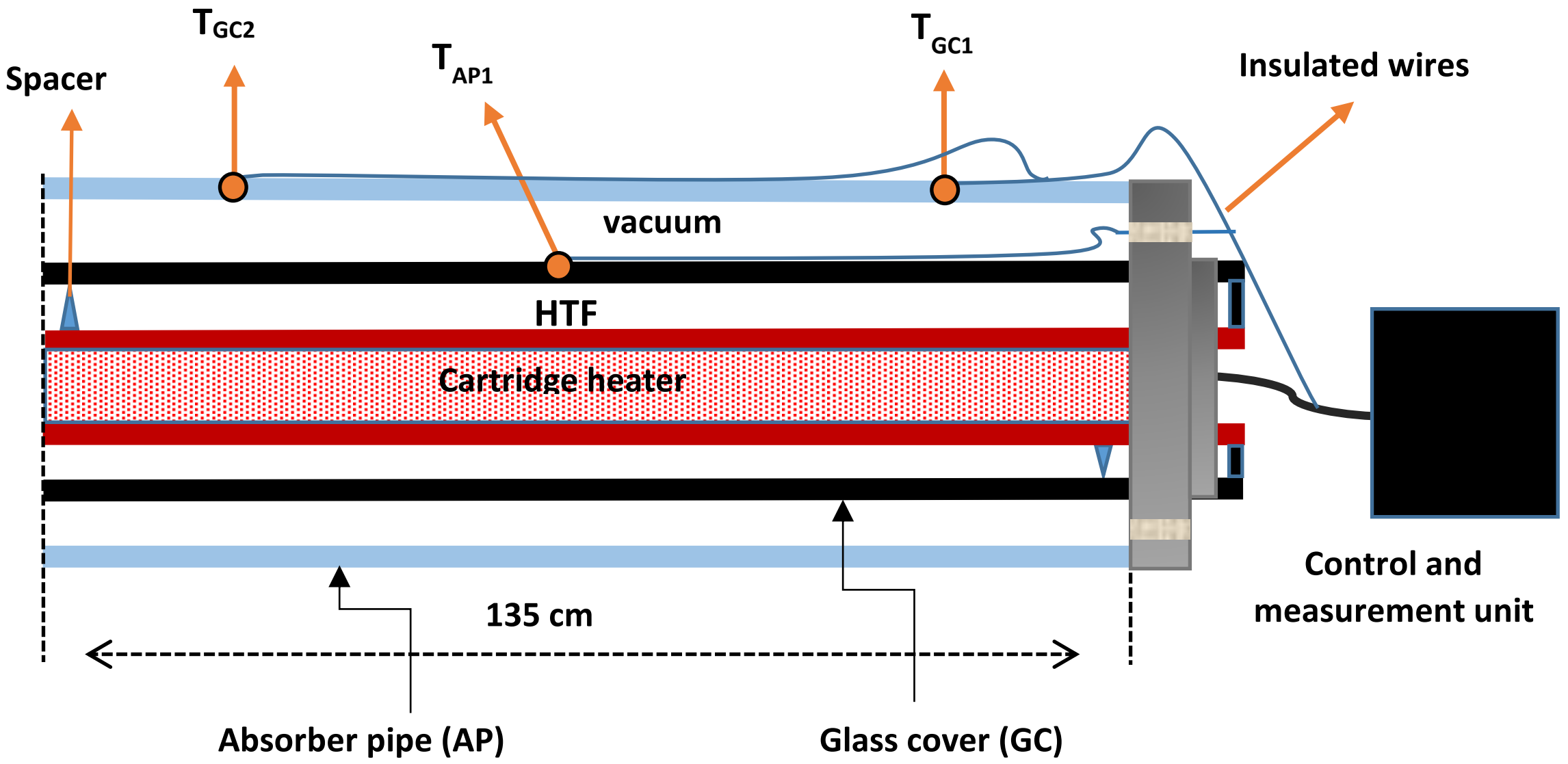

**Figure 3:** The right one-half of RU. T represents the thermocouple position.

## 3. Theory and simulation study

The source of the thermal energy is an electrical resistance heater wire with a constant rate of heat generation and can be modified by a variac. Fig. 3 shows the total heat transfer of the system and the interactions between its components. The physical basis of our model starts with a comprehensive description of the thermal interaction. Under steady operating conditions, the absorber pipe and the glass cover reach a different stagnation temperature. In addition, the heat loss and the heat gain of each element in the RU must equal the total rate of heat generation of the heating elements $\dot{E}_{gen}$

$$\dot{q}_{GC,amb} = \dot{q}_{GC,cond} = \dot{q}_{AP,GC} = \dot{q}_{AP,cond} = \dot{E}_{gen},$$

where $\dot{q}_{GC,amb}$ is the rate of the heat transfer from the glass cover (GC) to the surroundings, $\dot{q}_{GC,cond}$ is the conduction through the GC layer, $\dot{q}_{AP,GC}$ is the heat transfer from AP to GC, and $\dot{q}_{AP,cond}$ is the conduction through the AP.

The calculations start from the heat loss to the ambient because the ambient temperature is always known. We initially guess the unknown outer GC surface temperature $T_{g,o}$ iteratively, until the steady operating condition at which $\dot{q}_{GC,amb} = \dot{E}_{gen}$ is fulfilled. The heat rate $\dot{q}_{GC,amb}$ consists of natural convection and radiation heat transfer from the glass cover to the ambient.

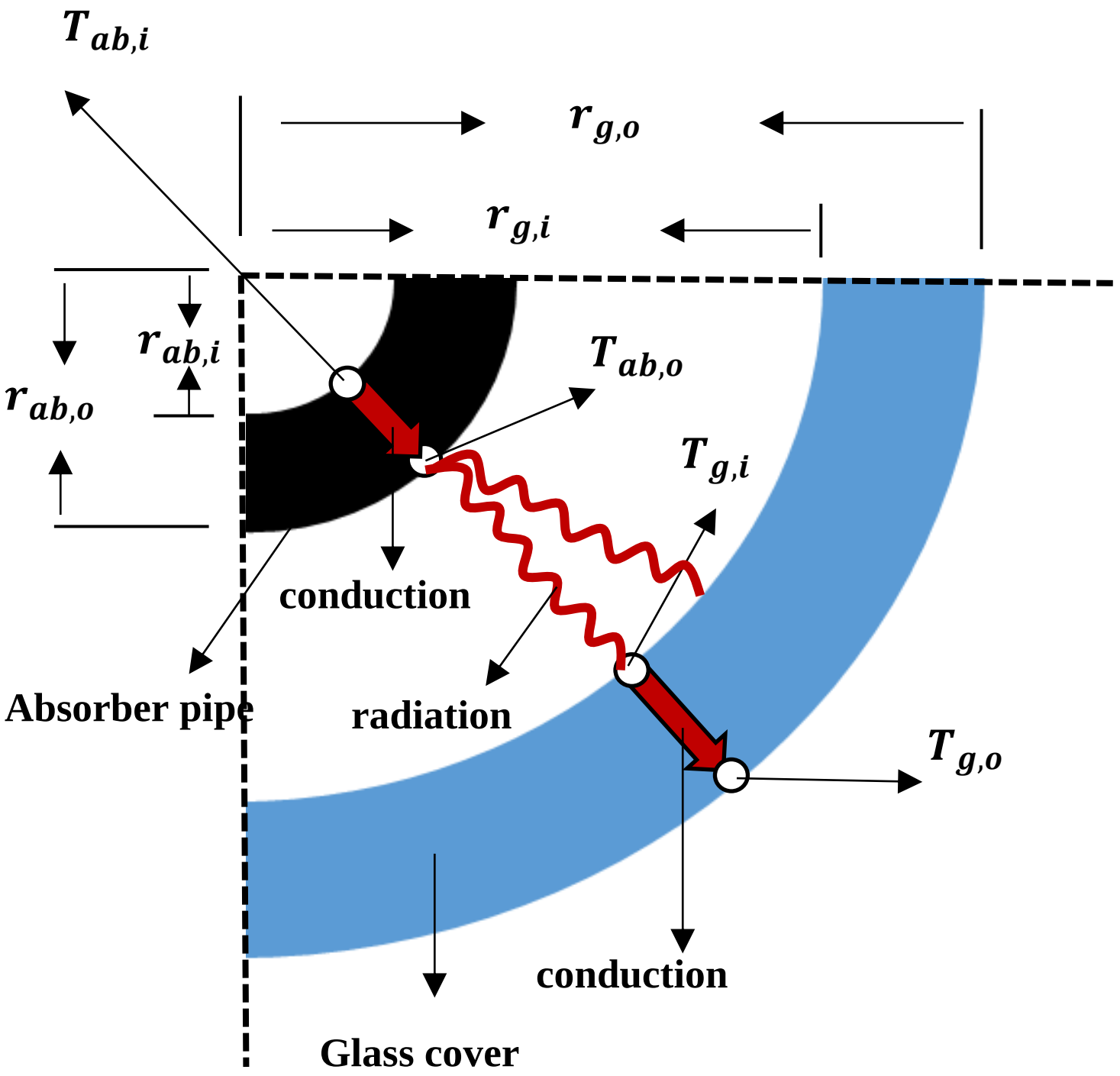

**Figure 4**: Receiver unit cross section.

The air properties during the calculation were selected at $T_{avg} = \frac{T_{g,o}+T_{amb}}{2}$. We could then evaluate $T_{g,i}$ at which the rate of heat loss due to the conduction through GC equal $\dot{E}_{gen}$. In the same way, $T_{ab,o}$ is evaluated through iteration until fulfilling $\dot{q}_{AP,GC} = \dot{E}_{gen}$, where $\dot{q}_{AP,GC}$ consists of the rate of the heat transfer between the AP and GC by convection and radiation. The convection heat transfer inside the evacuated annulus was ignored. Finally, $T_{ab,i}$ was evaluated such that the conduction heat loss through the AP equaled $\dot{E}_{gen}$. This simulation code was implemented using Python and is shown in Fig. 5.

## 4. Results

The initial experimental results tested for a normal RU reference unit without any coating. The aim was to validate our theoretical framework and simulation using experimental results. Two experiments were performed testing the thermal behaviour of the RU; 1) with air inside the annulus, and 2) with the annulus evacuated. The simulation predicted differing temperature readings for them, which we wanted to verify experimentally. The tested RU is shown in Fig. 2

In Fig. 6 and 7, the heat loss per meter of the RU length is depicted as a function of the absorber pipe and glass cover temperatures respectively. In the figures, "Sim" indicates simulation and "Exp" indicates experimental. In Fig. 6, the discrepancy between the simulation and experimental work in both cases vacuum and air in the annulus are 2% and 5.6% respectively.

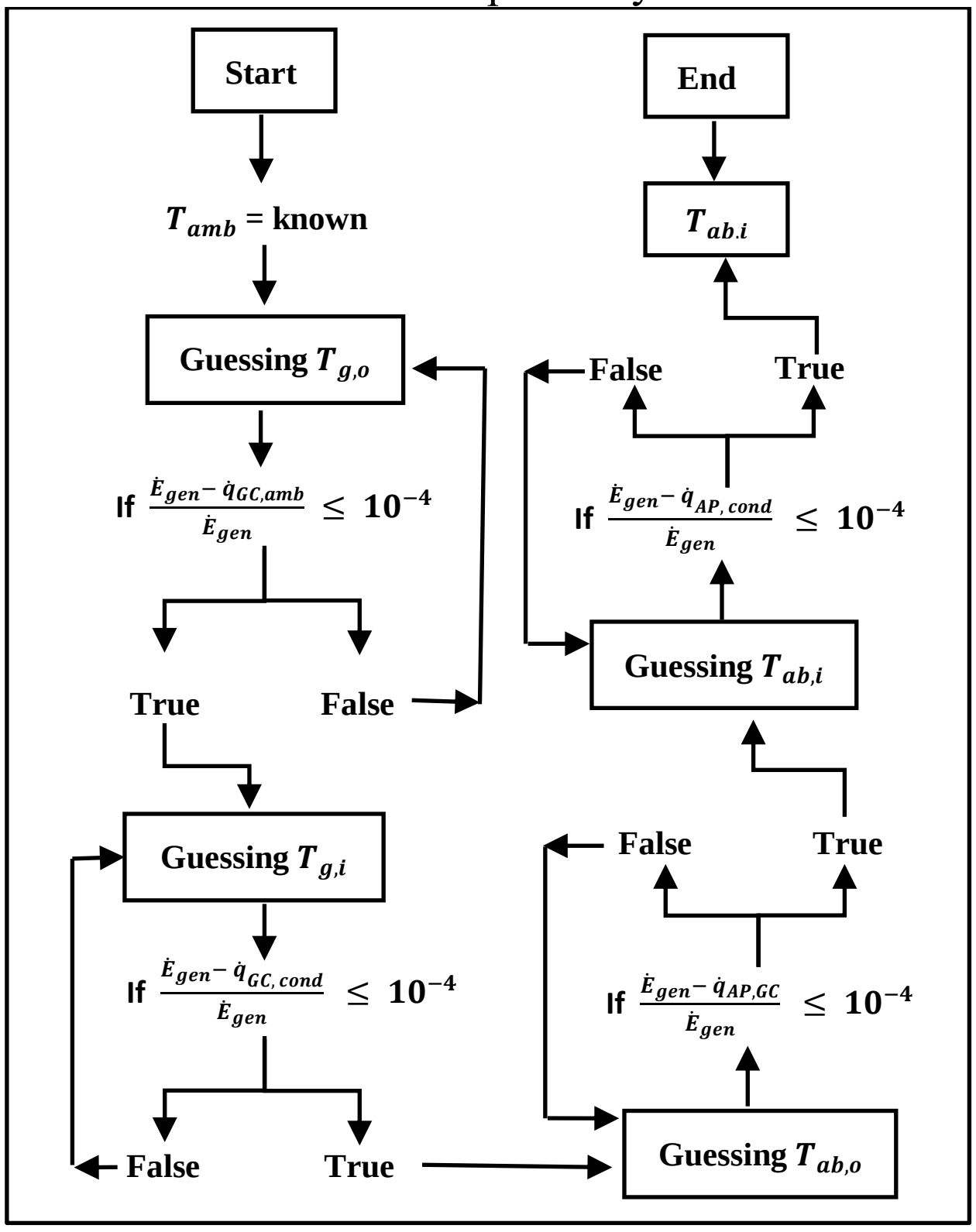

**Figure 5**: Algorithm for simulation code

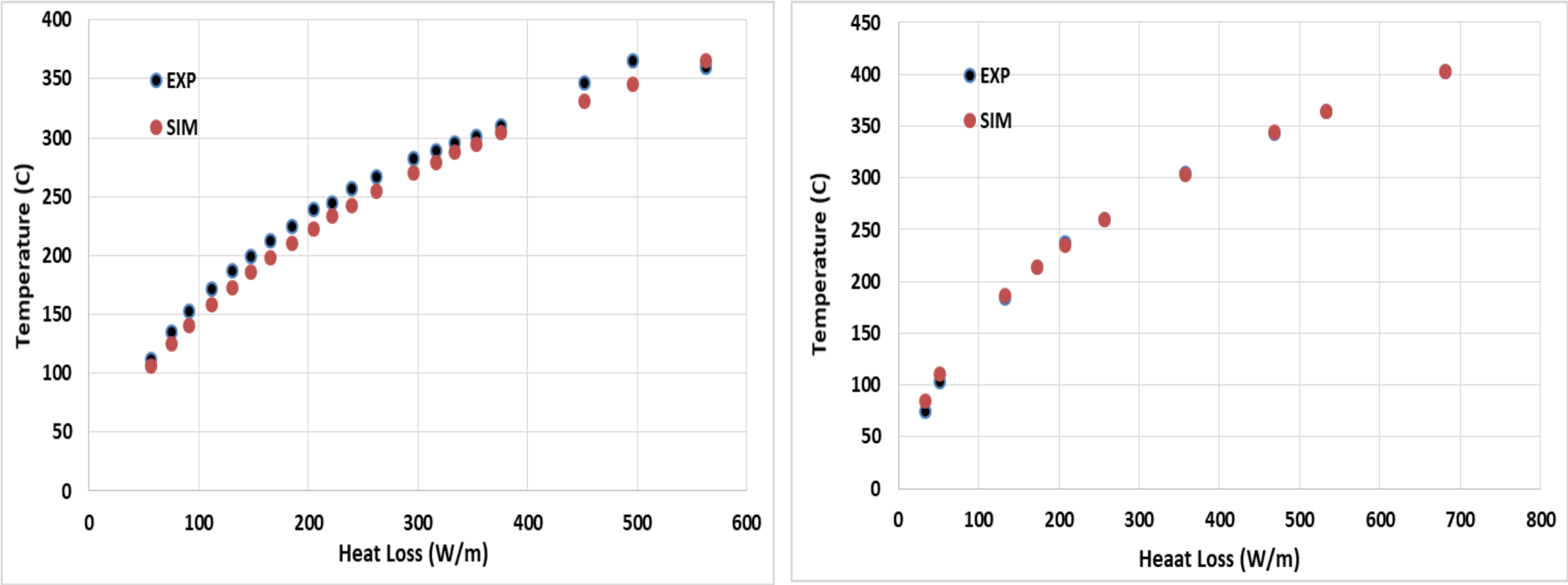

**Figure 6**: a) Heat loss versus AP temperature (Air in the annulus).

b) Heat loss versus AP temperature (Vacuum in the annulus).

The simulation predicted a higher AP temperature for the vacuum system, since there the convective heat transfer has been eliminated, and the only for heat to escape is via radiation. This is clearly shown in the experimental results, where the AP temperature is measurably higher in the vacuum case.

In all cases the simulation seems to predict a greater heat loss. This seems to suggest that some of the material property values need to be re-evaluated. Further, the edge heat loss effects need to be incorporated.

In Fig. 7, the simulation predicts that the GC temperature for the case of vacuum and air in the annulus is the same at a fixed value of the heat power generation, since the heat dissipation mechanism from the GC to the ambient is similar. Further, the discrepancy between the simulation and experimental work in both cases vacuum and air in the annulus are 6% and 3% respectively.

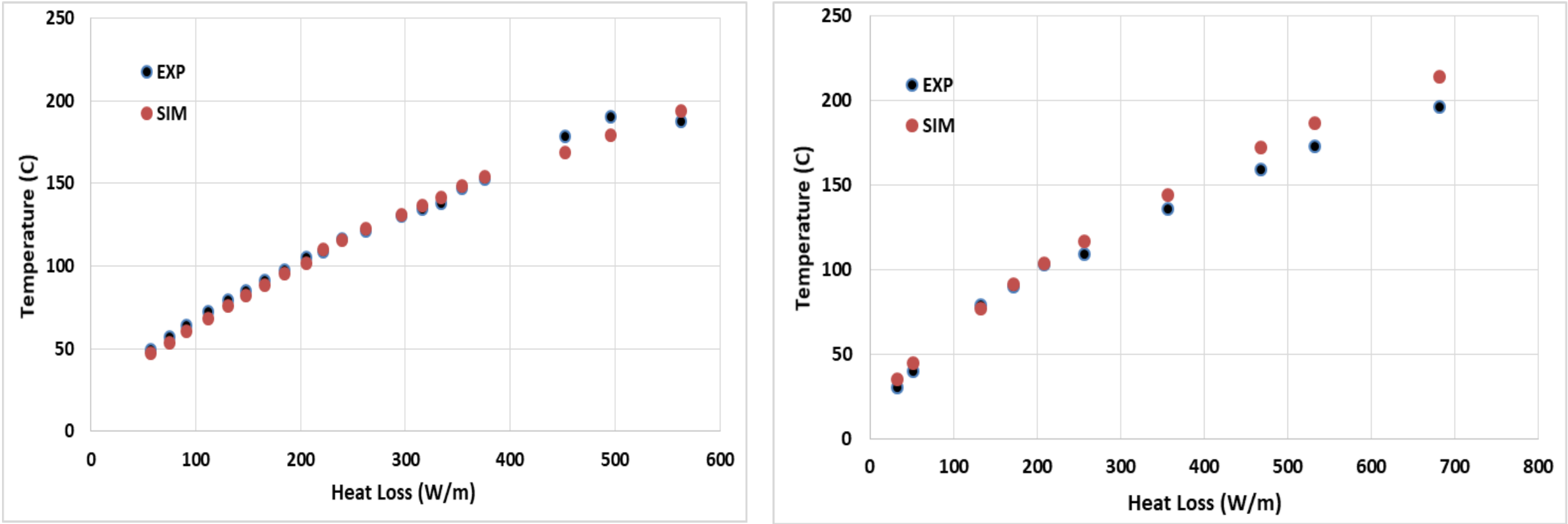

**Figure 7**: a) Heat loss versus GC temperature (Air in the annulus). b) Heat loss versus GC temperature (Vacuum in the annulus).

## 5. Conclusion

We tested a receiver unit (RU) of the parabolic trough collector indoor, in order to validate our theoretical and numerical framework. Two experiments were performed, for which the simulation predicted different results. Experimental data verified the results of the simulation to within 2 % and 6% discrepancy for the absorber pipe temperature in the case of vacuum and air in the annulus respectively. Also, a discrepancy of 6% and 3% for the glass cover in the case of vacuum and air in the annulus respectively. This successful set of validations encourages us to continue testing our proposed designs using this experimental setup.

## References

[1] Cyulinyana M C and Ferrer P 2011 Heat efficiency of a solar trough receiver with a hot mirror compared to a selective coating *South Afr. J. Sci.* **107** 01–7

[2] Sargent and Lundy LLC Consulting Group 2003 *Assessment of Parabolic Trough and Power Tower Solar Technology Cost and Performance Forecasts* (Chicago, Illinois: DIANE Publishing)

[3] Lampert C M 1979 Coatings for enhanced photothermal energy collection I. Selective absorbers *Sol. Energy Mater.* **1** 319–41

[4] Burkholder F and Kutscher C 2008 *Heat-loss testing of Solel's UVAC3 parabolic trough receiver* (United States: National Renewable Energy Laboratory (NREL), Golden, CO.)

[5] Kennedy C E and Price H 2005 Progress in Development of High-Temperature Solar-Selective Coating *ASME Int. Sol. Energy Conf. Sol. Energy* 749–55

[6] Twidell J and Weir T 2015 *Renewable Energy Resources* (New York: Routledge)

[7] Kreith F and Kreider J F 1978 *Principles of solar engineering* (United States: Hemisphere Pub. Corp.)

[8] Cyulinyana M C and Ferrer P 2011 Heat efficiency of a solar trough receiver with a hot mirror compared to a selective coating *South Afr. J. Sci.* **107** 01–07

[9] Liu D-S, Sheu C-S, Lee C-T and Lin C-H 2008 Thermal stability of indium tin oxide thin films co-sputtered with zinc oxide *Thin Solid Films* **516** 3196–3203
[10] William S J 2000 Engineering heat transfer *CRC Press Boca Raton Lond. N. Y. DC*
[11] Mohamad K and Ferrer P Computational comparison of a novel cavity absorber for parabolic trough solar concentrators *Submitt. Proc. 62th Annu. Conf. South Afr. Inst. Phys. SAIP2017*